\documentclass[12pt,a4paper]{iopart}

\usepackage[english]{babel}

\usepackage[utf8]{inputenc}
\usepackage[T1]{fontenc}

\usepackage{graphicx}

\usepackage{amsfonts}

\renewcommand{\vec}[1]{\bi{#1}}

\newcommand{\Ra}{\mathop{\mathrm{Ra}}}
\newcommand{\Nu}{\mathop{\mathrm{Nu}}}

\newcommand{\avg}[1]{\langle {#1} \rangle}
\newcommand{\condavg}[2]{\langle {#1} | {#2} \rangle}

\renewcommand{\Bigl}{\Big(}

\renewcommand{\Bigr}{\Big)}

\eqnobysec

\begin{document}

\title{Temperature Statistics in Turbulent Rayleigh-B\'enard Convection}

\author{J. L{\"u}lff, M. Wilczek, R. Friedrich}

\address{Institute for Theoretical Physics, University of M{\"u}nster, Wilhelm-Klemm-Str. 9, 48149 M{\"u}nster, Germany}

\ead{johannes.luelff@uni-muenster.de}

\begin{abstract}
  Rayleigh-B\'enard convection in the turbulent regime is studied using statistical methods.
  Exact evolution equations for the probability density function of temperature and velocity are derived from first principles within the framework of the Lundgren-Monin-Novikov hierarchy known from homogeneous isotropic turbulence.
  The unclosed terms arising in the form of conditional averages are estimated from direct numerical simulations.
  Focusing on the statistics of temperature, the theoretical framework allows to interpret  the statistical results in an illustrative manner, giving deeper insight into the connection between dynamics and statistics of Rayleigh-B\'enard convection.
  The results are discussed in terms of typical flow features and the relation to the heat transfer.
\end{abstract}

% 47.27.te 	Turbulent convective heat transfer
% 47.27.eb 	Statistical theories and models
% 47.20.Bp 	Buoyancy-driven instabilities (e.g., Rayleigh-Benard)
% 47.27.ek 	Direct numerical simulations

\submitto{\NJP}

\section{Introduction}
Rayleigh-B\'enard convection is a paradigm of a pattern forming system far from equilibrium.
Convective fluid motion in a vessel is induced by a vertical temperature gradient between the bottom and top boundaries due to buoyancy forces.
In dependence on this temperature gradient, the geometry of the experiment and the fluid properties, a whole zoo of instabilities has been observed ranging from laminar, spatially coherent convective motion over spatially ordered but temporally chaotic up to highly turbulent fluid motion.
We refer the reader to reviews available on the topic \cite{busse78rpp,bodenschatz00rfm}.

Recently, much efforts have been devoted to the analysis of turbulent Rayleigh-B\'enard (RB) convection both by experimental as well as theoretical means \cite{siggia94rfm,ahlers09rmp}.
Direct numerical simulations allow one to consider the dynamical and statistical properties of RB turbulence and the transitions between different types of flows in fine detail.

\begin{figure}[hb]
  \centering
  \includegraphics[width=0.6\textwidth]{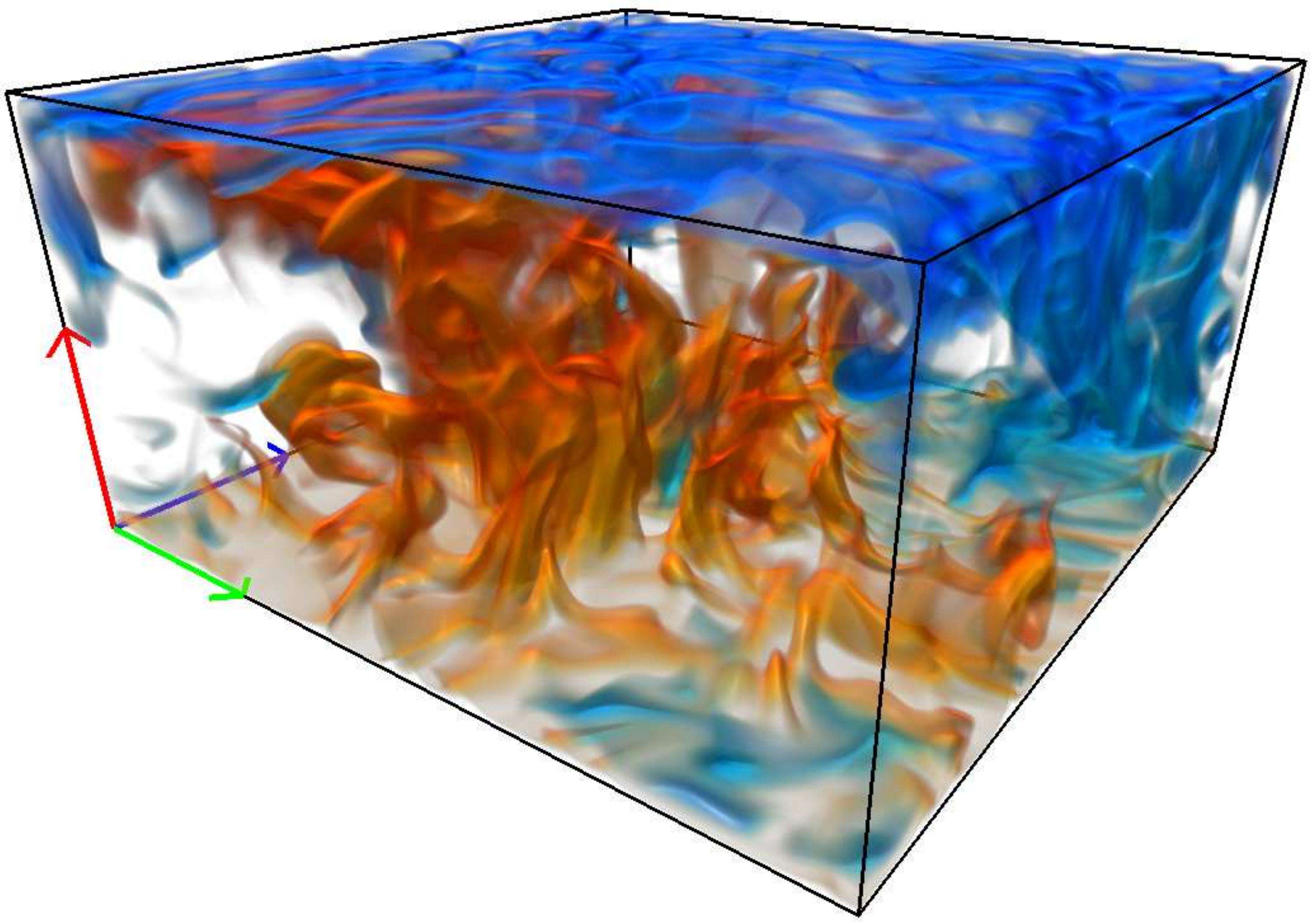}
  \caption{Snapshot of the temperature field $T(\vec{r})$. The green, blue and red arrows point in $x$-, $y$- and $z$-direction, respectively. The volume rendering has been done with the open source renderer \emph{Voreen} \cite{meyerspradow09iee}.}
  \label{fig:T_field_voreen}
\end{figure}

It is obvious that the analysis of turbulent convective fluid motion has to be based on a combination of tools from dynamical systems theory, statistical physics, and the theory of stochastic processes.
A necessary step is the statistical formulation of the underlying basic fluid dynamic equations, which for the most simple case are the Oberbeck-Boussinesq equations for the velocity field $\vec{u}(\vec{r},t)$, the temperature field $T(\vec{r},t)$, and the pressure field $p(\vec{r},t)$:
\begin{eqnarray}\label{eq:boussinesq}
  \left[\frac{\partial }{\partial t}+{\vec u}\cdot \nabla \right]T(\vec{r},t) &=& \Delta T({\vec r},t) \nonumber\\
  \left[\frac{\partial }{\partial t}+\vec{u}\cdot\nabla\right]\vec{u}(\vec{r},t) &=& -\nabla p(\vec{r},t)+\Pr \left[\Delta \vec{u}(\vec{r},t)+\Ra T(\vec{r},t) \vec{e}_z \right] \\
  \nabla \cdot {\vec u}({\vec r},t) &=& 0 \nonumber
\end{eqnarray}
The equations have been nondimensionalized using the Rayleigh number $\Ra=\frac{\alpha g \Delta T h^3}{\nu\kappa}$, which is a dimensionless measure of the temperature gradient across the fluid layer (with thermal expansion coefficient $\alpha$, gravitational acceleration $g$, outer temperature difference $\Delta T$, and distance of top and bottom plate $h$), as well as the Prandtl number $\Pr=\frac{\nu}{\kappa}$ as the ratio of kinematic viscosity $\nu$ to heat conductivity $\kappa$ of the fluid.
Thus, the vertical spatial coordinate obeys $z\in[0,1]$, and the boundary conditions of the temperature at the bottom and top plate are $T(z=0)=\frac{1}{2}$ and $T(z=1)=-\frac{1}{2}$.
For the velocity, no-slip boundary conditions $\vec{u}(z=0)=\vec{u}(z=1)=0$ are assumed.
These equations are solved numerically by a suitably designed penalization approach, described in \sref{sec:numerical_results}.
A snapshot of the temperature field is exhibited in \fref{fig:T_field_voreen}.

The statistical analysis is based on joint probability density functions (PDFs) for the temperature and the velocity at a single point in space and time.
The basic fluid dynamic equations require the validity of certain relations among these PDFs.
From these relations, corresponding expressions relating the various moments of the fields can be derived.
For the case of incompressible turbulence, these relations have been formulated by Lundgren, Monin and Novikov \cite{lundgren67pof,monin67pmm,novikov68sdp}, and Ulinich and Lyubimov \cite{ulinich68spj} and are sometimes known as the Lundgren-Monin-Novikov (LMN) hierarchy.
They are directly related to Hopf's functional equation, which can be viewed as the basic statistical formulation of the Navier-Stokes equation in the Eulerian framework \cite{monin07book}.
Similar relations can be derived for the corresponding Lagrangian quantities \cite{friedrich03prl}.
It is evident that an analogous treatment is feasible for the Oberbeck-Boussinesq equations.

In the present article we will use this approach in order to analyze the single-point temperature probability density function for stationary turbulent RB convection.
Our analysis combines direct numerical simulations with the relation of the LMN hierarchy for the single-point PDF.
The result is a partial differential equation for the temperature PDF.
The derivation of this relation is outlined in section \sref{sec:single_point_pdf}.

The aim is to formulate an equation which characterizes this PDF.
Starting from the Oberbeck-Boussinesq equations, we derive an evolution equation for the single-point joint probability density function of velocity and temperature along the lines of Lundgren, Monin and Novikov \cite{lundgren67pof,monin67pmm,novikov68sdp} for the case of incompressible turbulence.
This equation is unclosed due to the fact that it contains unclosed expressions which can be related to fluid pressure, viscous dissipation and heat diffusion.
However, these expressions can be treated by introducing conditional averages, which can be extracted from direct numerical simulations. This leads to a partial differential equation for the joint temperature-velocity PDF.
A similar approach has been performed by Novikov \cite{novikov93fdr,novikov94mpl}, and more recently by Wilczek et al.~for the PDFs of vorticity \cite{wilczek09pre} and velocity \cite{wilczek10} for stationary, isotropic turbulence.
On the other hand, modeling of unclosed terms is also a possible method, as performed by e.g.~Pope \cite{pope81pof,pope00book}.
As we shall indicate, the analysis of the evolution equation for the temperature PDF yields a comprehensive description of the dynamical processes in RB convection.

The article is structured as follows:
In \sref{sec:single_point_pdf} we will derive the evolution equation for the temperature-velocity joint PDF.
In \sref{sec:single_point_temperature_pdf}, we reduce the joint PDF to the temperature PDF and make use of statistical symmetries to cut down the complexity of the evolution equation.
Then we present a descriptive way to deal with this equation involving the method of characteristics.
\Sref{sec:heat_transport} briefly discusses connections to the Nusselt number, relevant for the heat transport.
These theoretical results are complemented by results from direct numerical simulations, which will be discussed in \sref{sec:numerical_results}, followed by a summary in \sref{sec:summary}.

\section{Single-Point PDF}
\label{sec:single_point_pdf}
We are interested in the joint temperature-velocity probability distribution $f(\tau,\vec{v};\vec{r},t)$ and want to derive the corresponding evolution equation.
Formally, the probability density function is obtained as a suitable average over the so-called fine-grained probability distribution
\begin{equation}\label{eq:fine_grained_pdf}
  \widehat{f}(\tau,\vec{v};\vec{r},t)=\delta(\tau-T(\vec{r},t))\: \delta(\vec{v}-\vec{u}(\vec{r},t)).
\end{equation}
It is important to distinguish between \emph{sample space} variables $\tau$, $\vec{v}$ and the corresponding \emph{realizations} of the temperature and velocity fields $T(\vec{r},t)$, $\vec{u}(\vec{r},t)$.
Therefore, one could think of $\widehat{f}$ as the PDF of one particular realization of the fields.
Also, the notation of the arguments in $\widehat{f}(\tau,\vec{v};\vec{r},t)$ emphasizes the difference between the sample space variables $\tau$, $\vec{v}$ and the coordinates $\vec{r}$, $t$ -- the PDF is normalized with respect to the sample space variables, the coordinates are just parameters.

The full probability density function is now obtained as an ensemble average over all possible realizations of the temperature and velocity fields:
\begin{equation}\label{eq:full_pdf}
  f(\tau,\vec{v};\vec{r},t) = \avg{\widehat{f}(\tau,\vec{v};\vec{r},t)} = \avg{\delta(\tau-T(\vec{r},t))\:\delta(\vec{v}-\vec{u}(\vec{r},t))}
\end{equation}
The brackets $\avg{\cdot}$ denote the ensemble average, in contrast to spatial averages $\avg{\cdot}_V$ and $\avg{\cdot}_A$ over the whole fluid volume, or a horizontal plane at height $z$, respectively.

The definition of the fine-grained PDF \eref{eq:fine_grained_pdf} can be differentiated with respect to the space and time variables, giving
\begin{equation}\label{eq:gradient_fine_grained_pdf}
  \frac{\partial }{\partial x_i} \widehat{f} = -\left[\frac{\partial}{\partial \tau} \frac{\partial T(\vec{r},t)}{\partial x_i}\widehat{f} + \nabla_\vec{v}\cdot\frac{\partial\vec{u}(\vec{r},t)}{\partial x_i}\widehat{f}\right]
\end{equation}
as the spatial gradient, and an analogous equation for the temporal derivative $\frac{\partial }{\partial t} \widehat{f}$.
Note that the operators $\frac{\partial}{\partial\tau}$ and $\nabla_\vec{v}$ act on $\widehat{f}$.
Now, multiplying \eref{eq:gradient_fine_grained_pdf} by $u_i$ and adding the temporal derivative allows us to make use of the basic Oberbeck-Boussinesq equations \eref{eq:boussinesq}.
This results in the desired evolution equation for the fine-grained PDF:
\begin{eqnarray}
  \frac{\partial }{\partial t} \widehat{f} + \vec{u}\cdot\nabla\widehat{f} &=& -\left[\frac{\partial}{\partial\tau}\left(\frac{\partial T}{\partial t}+\vec{u}\cdot\nabla T\right)\widehat{f} + \nabla_\vec{v}\cdot\left(\frac{\partial \vec{u}}{\partial t}+\vec{u}\cdot\nabla\vec{u}\right)\widehat{f}\right] \nonumber\\
  &=& -\frac{\partial}{\partial \tau} \Delta T \widehat{f} -\nabla_\vec{v}\cdot[ -\nabla p+\Pr\left( \Delta \vec{u} + \Ra T\vec{e}_z \right)] \widehat{f}
\end{eqnarray}

Performing the ensemble average of this equation in order to arrive at an equation for the full PDF \eref{eq:full_pdf}, one encounters the closure problem of turbulence, since the unclosed averages $\avg{\Delta T \widehat{f}}$, $\avg{-\nabla p \widehat{f}}$ and $\avg{\Delta \vec{u} \widehat{f}}$ show up.
The LMN \emph{hierarchy} ansatz would mean to treat these terms via a coupling to the two-point PDF; the evolution equation of this two-point PDF would in turn introduce a coupling to the three-point PDF, and so on \cite{lundgren67pof}.

Instead of introducing this hierarchy of coupled evolution equations for the multi-point PDFs, our strategy \cite{wilczek09pre,wilczek10} is to express the unclosed terms as conditional averages, since these are accessible to direct numerical simulations.
The result is a partial differential equation governing the joint temperature-velocity PDF and relating its shape as a function of space point $\vec{r}$ to the conditional averages.
The functional form of these conditional averages is a signature of the underlying dynamical processes of RB convection.

Introducing the conditional averages
\begin{eqnarray}
 \label{eq:cond_avg}
 \avg{\Delta T \widehat{f}} &=& \condavg{\Delta T}{\tau,\vec{v},\vec{r},t} f \nonumber\\
 \avg{-\nabla p \widehat{f}} &=& \condavg{-\nabla p}{\tau,\vec{v},\vec{r},t} f \\
 \avg{\Delta \vec{u} \widehat{f}} &=& \condavg{\Delta \vec{u}}{\tau,\vec{v},\vec{r},t} f \nonumber
\end{eqnarray}
we arrive at the following relation for the single-point probability distribution $f(\tau,\vec{v};\vec{r},t)$:

\begin{eqnarray}\label{eq:eqn_of_motion_of_jpdf_1}
  \fl\frac{\partial}{\partial t}f + \vec{v} \cdot \nabla f = &-&\frac{\partial}{\partial \tau}\condavg{\Delta T}{\tau,\vec{v},\vec{r},t} f \nonumber \\
  &-& \nabla_\vec{v}\cdot\left[\condavg{-\nabla p}{\tau,\vec{v},\vec{r},t} + \Pr \left( \condavg{\Delta \vec u}{\tau,\vec{v},\vec{r},t} + \Ra\tau\vec{e}_z \right)\right]f
\end{eqnarray}

Since at the boundaries of the RB cell the velocity and temperature fields are statistically sharp quantities, the probability distribution has to obey the conditions $f(\tau,\vec{v};z=0)=\delta\left(\tau-\case{1}{2}\right)\delta(\vec{v})$ and $f(\tau,\vec{v};z=1)=\delta\left(\tau+\case{1}{2}\right)\delta(\vec{v})$ for arbitrary $x,y$.

We note in passing that a different version of the evolution equation can be obtained by introducing the Laplacian of the PDF. 
Specializing in the case $\Pr=1$, it is possible to re-express the conditional averages $\condavg{\Delta T}{\tau,\vec{v},\vec{r},t}$ and $\condavg{\Delta \vec{u}}{\tau,\vec{v},\vec{r},t}$ via the relation
\begin{eqnarray} \label{eq:laplace_f}
  \fl\Delta f = &-&\frac{\partial}{\partial \tau} \avg{\Delta T \widehat{f}} - \nabla_\vec{v}\cdot\avg{\Delta \vec{u} \widehat{f}} \nonumber \\
  \fl &+& \frac{\partial^2}{\partial \tau^2} \avg{(\nabla T)^2 \widehat{f}} + 2 \frac{\partial^2}{\partial \tau \partial v_j} \avg{\nabla T \cdot \nabla u_j \widehat{f}}  + \frac{\partial^2}{\partial v_i \partial v_j} \avg{\nabla u_i \cdot \nabla u_j \widehat{f}}.
\end{eqnarray}
Here and in the following, Einstein's summation convention over repeated indices is used.
Again, we can introduce conditional expectations, where the arguments of the conditional averages have been abbreviated as $\star\,\widehat{=}\,\tau,\vec{v},\vec{r},t$.
The resulting evolution equation reads
\begin{eqnarray}\label{eq:eqn_of_motion_of_jpdf_2}
  \fl \frac{\partial}{\partial t}f+\vec{v} \cdot \nabla f = \Delta f &-& \nabla_\vec{v}\cdot \left[\condavg{-\nabla p}{\star} + \Ra\tau \vec{e}_z \right]f - \frac{\partial^2}{\partial \tau^2}\condavg{(\nabla T)^2}{\star} f \nonumber \\
  \fl &-& 2\frac{\partial^2}{\partial \tau \partial v_j} \condavg{\nabla T \cdot \nabla u_j }{\star}f -  \frac{\partial^2}{\partial v_i \partial v_j} \condavg{\nabla u_i \cdot \nabla u_j}{\star}f.
\end{eqnarray}
A somewhat more complicated equation holds for $\Pr \ne 1$.
This relationship shows that the single-point joint temperature-velocity PDF $f(\tau,\vec{v};\vec{r},t)$ is essentially determined by the conditionally averaged dissipation-like terms $\condavg{(\nabla T)^2}{\star}$, $\condavg{\nabla T \cdot \nabla u_j}{\star}$ and $\condavg{\nabla u_i \cdot \nabla u_j}{\star}$ as well as the conditional pressure gradient $\condavg{-\nabla p}{\star}$.

\section{Single-Point Temperature PDF and Implications of Statistical Symmetries}\label{sec:single_point_temperature_pdf}
In the following we shall restrict our attention to the reduced temperature probability distribution and its evolution equation.
As it will turn out, already this equation gives insight into the connection between RB dynamics and temperature statistics, besides obviously describing the temperature statistics itself.
Also, because the final evolution equation of temperature PDF involves scalar valued functions only, a numerical approach is easily feasible.

The reduced temperature PDF is obtained by integrating out the velocity part: 
\begin{equation}
  h(\tau;\vec{r},t)=\int\! \rmd^3v\, f(\tau,\vec{v};\vec{r},t)
\end{equation}
Starting from equation \eref{eq:eqn_of_motion_of_jpdf_1} we obtain the simple equation
\begin{equation}\label{eq:eqn_of_motion_of_temperature_pdf_1}
  \frac{\partial}{\partial t}h + \nabla \cdot \condavg{\vec{u}}{\tau,\vec{r},t}h = -\frac{\partial }{\partial \tau} \condavg{\Delta T}{\tau,\vec{r},t} h,
\end{equation}
where we have performed the integration with respect to the velocity $\vec{v}$.
Thereby, we had to introduce the conditional averages $\condavg{\vec{u}}{\tau,\vec{r},t}$ and $\condavg{\Delta T}{\tau,\vec{r},t}$.
Alternatively, one could re-enact the derivation performed in \sref{sec:single_point_pdf} for the fine-grained PDF of the temperature $\widehat{h}(\tau;\vec{r},t)=\delta(\tau-T(\vec{r},t))$.

Analogous to \eref{eq:eqn_of_motion_of_jpdf_2}, we can derive a further relation using the identity
\begin{equation}
  \Delta h = \avg{\Delta\widehat{h}} = -\frac{\partial}{\partial \tau} \avg{\Delta T(\vec{r},t)\:\widehat{h}} + \frac{\partial^2}{\partial \tau^2} \avg{\left(\nabla T(\vec{r},t) \right)^2\widehat{h}}.
\end{equation}
This relation also follows from \eref{eq:laplace_f}.
With this equation, we can now summarize the equation for the temperature PDF $h(\tau;\vec{r},t)$ in the form
\begin{equation}\label{eq:eqn_of_motion_of_temperature_pdf_2}
  \frac{\partial}{\partial t}h + \nabla \cdot \condavg{\vec{u}}{\tau,\vec{r},t} h = \Delta h - \frac{\partial^2}{\partial \tau^2}\condavg{\left(\nabla T\right)^2}{\tau,\vec{r},t} h.
\end{equation}
Here, the conditional average of the term $(\nabla T)^2$ comes up, which is related to the Nusselt number.
Details will be discussed in \sref{sec:heat_transport}.

The evaluation of the conditional averages appearing in \eref{eq:eqn_of_motion_of_temperature_pdf_1} is greatly simplified by considering convection that is statistically stationary in time and has periodic horizontal boundaries (i.e.~is homogeneous with respect to the horizontal coordinates).
Under the assumption of a statistically stationary flow, the PDFs and therefore the conditional averages cannot depend on the time variable; also the dependency on the horizontal coordinates drops out.
So instead of dealing with statistical quantities that depend on $\vec{r}$ and $t$, we simply have to retain the $z$-dependence.

Let us first consider the determining equation for the temperature PDF $h(\tau;z)$ in the form
\begin{equation}\label{eq:pde_h}
  \frac{\partial}{\partial z} \condavg{u_z}{\tau,z} h(\tau;z) + \frac{\partial}{\partial \tau}  \condavg{\Delta T}{\tau,z} h(\tau;z)=0,
\end{equation}
obtained from \eref{eq:eqn_of_motion_of_temperature_pdf_1}.
This equation in principle has to be solved together with the boundary conditions $h(\tau;0)=\delta\Bigl\tau-\frac{1}{2}\Bigr$ and $h(\tau;1)=\delta\Bigl\tau+\frac{1}{2}\Bigr$ and with appropriately modeled expressions for the conditional averages $\condavg{u_z}{\tau,z}$ and $\condavg{\Delta T}{\tau,z}$.
This direct approach will not be conducted in the present paper though, but can be taken as a starting point for future modeling.

Instead, this first order partial differential equation (PDE) can be analyzed with the help of the method of characteristics \cite{courant62book}.
Applying this method, one can find curves $\Bigl\tau(s), z(s)\Bigr$ in the $\tau$-$z$--phase space parameterized by $s$ along which the PDE \eref{eq:pde_h} transforms into an ordinary differential equation which can be integrated.
This approach will be sketched in the following.

Writing $h(s)=h\Bigl\tau(s);z(s)\Bigr$ and calculating the derivative of it gives
\begin{equation}
  \frac{\rmd}{\rmd s}h\Bigl\tau(s), z(s)\Bigr = \frac{\partial h}{\partial\tau}\frac{\rmd\tau}{\rmd s} + \frac{\partial h}{\partial z}\frac{\rmd z}{\rmd s}.
\end{equation}
The PDE \eref{eq:pde_h} is re-expressed in the form
\begin{equation}
  \condavg{\Delta T}{\tau,z}\frac{\partial h}{\partial\tau} + \condavg{u_z}{\tau,z}\frac{\partial h}{\partial z} = -\left( \frac{\partial}{\partial\tau}\condavg{\Delta T}{\tau,z} + \frac{\partial}{\partial z}\condavg{u_z}{\tau,z} \right) h.
\end{equation}
Comparing these two equations identifies the characteristic curves as solutions of
\begin{eqnarray}\label{eq:characteristics}
  \frac{\rmd}{\rmd s}\tau &=& \condavg{\Delta T}{\tau,z} \nonumber\\
  \frac{\rmd}{\rmd s}z &=& \condavg{u_z}{\tau,z}
\end{eqnarray}
Along these curves, the PDE \eref{eq:pde_h} becomes
\begin{equation}
  \frac{\rmd}{\rmd s}h = -\left(\frac{\partial}{\partial \tau}\condavg{\Delta T}{\tau,z} + \frac{\partial}{\partial z}\condavg{u_z}{\tau,z}\right)h,
\end{equation}
which can be integrated to
\begin{equation}\label{eq:temperature_pdf_integral}
  h(s) = h(s_0)\exp\left[-\int\limits_{s_0}^s\!\rmd s \left(\frac{\partial}{\partial \tau}\condavg{\Delta T}{\tau,z} + \frac{\partial}{\partial z}\condavg{u_z}{\tau,z}\right)\right].
\end{equation}
This equation describes the evolution of the PDF along a trajectory $\Bigl\tau(s),z(s)\Bigr$ starting at point $\Bigl\tau(s_0),z(s_0)\Bigr$ in phase space.
A particularly appealing property of this formalism is that it allows to interpret the statistical results in an illustrative manner, because the characteristics, i.e.~trajectories in $\tau$-$z$--phase space, show the evolution of the ``averaged'' physical process.

It is tempting to interpret the characteristics as a kind of Lagrangian dynamics of a tracer particle inside the RB cell.
However, the dynamic of a tracer particle is stochastic, whereas the characteristics defined by \eref{eq:characteristics} describe purely deterministic trajectories and, thus, take the stochastic properties into account only in an averaged way.
In a sense, the characteristics describe the averaged evolution of an ensemble of fluid particles that are defined by their initial condition in the $\tau$-$z$--plane.

Thinking of turbulent RB convection with some physical intuition, one can expect certain features from the statistical quantities introduced in this section.
The conditionally averaged vertical velocity $\condavg{u_z}{\tau,z}$ should show positive correlation with the temperature, i.e.~it should mirror the well-known fact that hot fluid rises up and cold fluid sinks down.
Also the no-slip boundaries should be recognizable for $z\approx0$ and $z\approx1$, respectively.
The absolute value of the heat diffusive term $\condavg{\Delta T}{\tau,z}$ should be highest near the boundaries because of the sharp change of the temperature profile.
As the characteristics in a way describe the average path a fluid particle takes through $\tau$-$z$--phase space, the typical Rayleigh-B\'enard cycle of fluid heating up at the bottom, rising up, cooling down at the top and sinking down again should find its correspondence in the statistical quantities describing the evolution of the PDF.
Actual numerical studies of these quantities will be discussed in detail in \sref{sec:numerical_results}.

In the 1990s, an approach similar to the presented one was undertaken by Yakhot and followers, where PDFs of various quantities in a stationary flow were expressed as integrals over conditionally averaged variables.
Considered quantities included passive scalars by Sinai and Yakhot \cite{sinai89prl}, active scalars, such as temperature fluctuations by Yakhot \cite{yakhot89prl} and temperature increments by Ching \cite{ching93prl}, and even general functions of an arbitrary quantity measured in the flow by Pope and Ching \cite{pope93pfa,ching96pre}.
The conditional averages included spatial and temporal derivatives of these quantities.

%Yet, what sets our approach apart from the aforementioned ones is that we do not assume homogeneity in the spatial coordinates, i.e.~we still have the $z$-dependency present in our PDF equations.
In contrast to these works we do not assume homogeneity in the spatial coordinates, i.e.~we still have the $z$-dependency present in our PDF equations.
This allows us to discuss the PDFs with respect to the $z$-coordinate and observe qualitatively different statistics in different regions of the flow.
As a result, we are able to see the differing behaviour of bulk and boundary parts of the convection cell and how these are connected to the conditional averages and their dependency on the vertical position.
This has not been addressed in the literature.

\section{Connection to the Heat Transport}\label{sec:heat_transport}
The Nusselt number $\Nu$ as the ratio of convective to conductive heat transfer plays a key role in the analysis of RB convection.
It serves as a measure of how efficient heat can be transported through the convection cell.

Though we did not consider the Nusselt number so far, we note that there is an interesting connection between $\Nu$ and the conditional averages that appear in our calculations.
Because the Nusselt number may be defined as the volume average $\Nu=\avg{(\nabla T)^2}_V$ in non-dimensional units, the following expression comes up:
\begin{equation}\label{eq:nusselt1}
  \Nu = \frac{1}{V}\int\! \rmd^3r\,\rmd\tau\,\condavg{\left(\nabla T\right)^2}{\tau,\vec{r}} h(\tau;\vec{r})
\end{equation}
Therefore, $\condavg{\left(\nabla T\right)^2}{\tau,\vec{r}}$ can be viewed as a conditional Nusselt number density.
We point out that this quantity should be of considerable interest for the evaluation of theories concerning the Rayleigh number dependency of the Nusselt number based on a decomposition of the heat transport into bulk and boundary contributions, which underlies the Grossmann-Lohse theory \cite{grossmann00jfm} outlined in the review of Ahlers et al.~\cite{ahlers09rmp}. Also, this term is linked with the temperature dissipation rate, which is discussed in \cite{emran08jfm} with respect to temperature PDFs.

In a similar manner, we can employ the temperature-velocity joint PDF $f(\tau,\vec{v};\vec{r})$ to derive an equation relating Rayleigh and Nusselt number.
From the relation $\Ra(\Nu-1)=\avg{(\nabla\vec{u})^2}_V$ it is straightforward to see that
\begin{equation}\label{eq:nusselt2}
  \fl\int\! \rmd^3r\, \rmd\tau\, \rmd^3v\, f(\tau,\vec{v};\vec{r})\left[ \Ra (\condavg{(\nabla T)^2}{\tau,\vec{v},\vec{r}}-1)-\condavg{(\nabla \vec{u})^2}{\tau,\vec{v},\vec{r}}\right]=0.
\end{equation}
These two exact relations underline the importance of the conditional averages of $(\nabla T)^2$ and $(\nabla\vec{u})^2$ that naturally come up in our derivations.

\section{Numerical Results}\label{sec:numerical_results}
The benefit of our theoretical approach is that we can easily provide it with measurements and data in the form of numerical results.
To this end, we solve the basic Oberbeck-Boussinesq equations \eref{eq:boussinesq} with a standard dealiased pseudospectral code on a three-dimensional equidistant Cartesian grid with periodic boundary conditions. For an introduction into this topic the reader is referred to \cite{boyd01book,canuto87book}.

Periodic boundaries are required in horizontal direction, but in vertical direction Dirichlet conditions for velocity and temperature (i.e.~no-slip boundaries of constant temperature) are needed.
They are enforced by a \emph{volume penalization} ansatz \cite{angot99num,schneider05caf,keetels07jcp}:
The fluid domain $\Omega=[0,L_x] \times [0,L_y] \times [0,1] \subset \mathbb{R}^3$ is embedded in a computational domain that is extended by a layer of thickness $d$ in $z$-direction, $\Omega_c=[0,L_x]\times[0,L_y]\times[-d,1+d]$.
Inside the fluid domain $\Omega\subset\Omega_c$ the unaltered Oberbeck-Boussinesq equations are solved, while in the appended extra regions $\Omega_c\backslash\Omega$ a strong exponential damping ($-\frac{1}{\eta}\vec{u}$ and $-\frac{1}{\eta}\theta$, respectively, with $\eta\ll 1$) is added to the evolution equations \eref{eq:boussinesq} of velocity and temperature that damps the fields to zero.
By simulating the deviation from the linear temperature profile, $\theta(\vec{r},t):=T(\vec{r},t) + \left(z - \frac{1}{2}\right)$, instead of the temperature itself, the desired boundary conditions read $\vec{u}=0$ and $\theta=0$ for $z=0$ and $z=1$.
This change of variables $T\rightarrow\theta$ allows us to make use of the volume penalization approach in a straightforward manner.

The reason we choose this numerical scheme instead of often used Chebyshev-based codes, described for example in \cite{boyd01book} and used in e.g.~\cite{clercx06caf,schumacher09pre}, is because it allows for almost arbitrary shaped boundaries and sidewalls.
Although this feature is not used in the present paper due to the required horizontal homogeneity, it even allows to simulate cylindrical vessels on a (numerically cheap) Cartesian grid.
A more detailed report on this will be published in the future.

Our theoretical derivation relies on the concept of ensemble averages.
Of course, through our numerics we can only access a finite subset of all possible ensemble members.
So due to the statistical symmetries and by assuming ergodicity, the ensemble average is  substituted for a combined volume--time average in the numerics.
Likewise, the volume-averages $\avg{\cdot}_V$ and $\avg{\cdot}_A$ introduced earlier are actually evaluated as combined volume and time averages.
Here, time averaging means averaging over $1250$ statistically independent snapshots of the fields.
The line plots below only show the parts of the statistical quantities where the statistics converged, i.e.~where a significant number of events was obtained.

The simulation was conducted for the parameters $\Ra=4.33\times10^7$ and $\Pr=1$.
The computational domain $\Omega_c$ is resolved with $N_x\times N_y\times N_z=256\times256\times192$ gridpoints on an equidistant Cartesian grid, where the fluid domain $\Omega\subset\Omega_c$ is represented by $256\times256\times128$ gridpoints.
Thus, the aspect ratio is $\Gamma=2$, with the two horizontal dimensions being identical.
The Nusselt number is estimated to $\Nu=24.2$. 
We mention that we repeated the analysis presented below with a simulation of aspect ratio $\Gamma=4$ and found basically the same features, due to the fact that for sufficiently large $\Gamma$ the periodic boundaries do not have a significant influence on the flow.

\begin{figure}[ht]
  \centering
  {\footnotesize(a)}\hspace{0.5\textwidth}{\footnotesize(b)}\newline
  \includegraphics[width=0.53\textwidth]{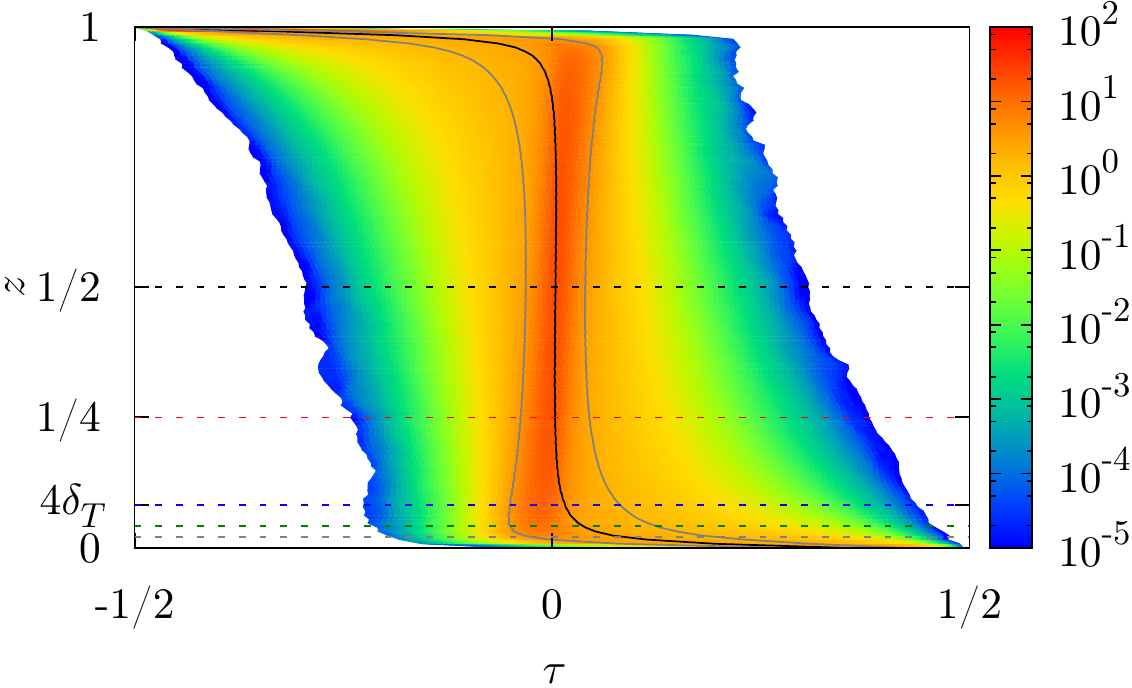}
  \hfill
  \includegraphics[width=0.45\textwidth]{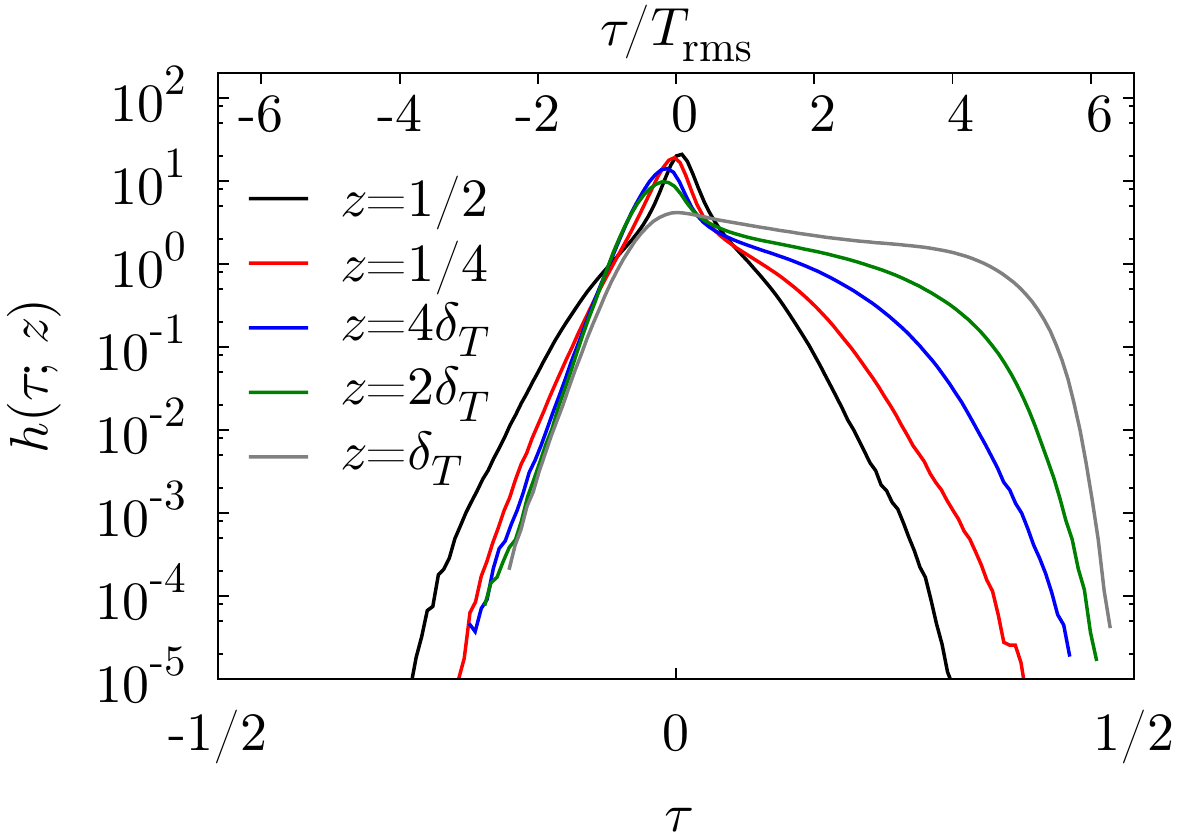}
  \caption{
    (a) The mean temperature $\avg{T(z)}_A$ and a color plot of the logarithm of the temperature PDF $h(\tau;z)$ as a function of $z$.
    The additional solid gray lines mark the contour line for $\avg{T(z)}_A\pm\big\langle\big(T-\avg{T(z)}_A\big)^2\big\rangle_A^{1/2}$ and indicate the standard deviation of temperature at height $z$.
    The horizontal dashed lines indicate the positions of slices in $\tau$-direction at fixed height and are located at $z\in\{\frac{1}{2},\frac{1}{4},4\delta_T,2\delta_T,\delta_T\}$, where $\delta_T=\frac{1}{2\Nu}$ is the thermal boundary layer thickness.
    (b) The logarithm of the temperature PDF $h(\tau;z)$ for different values of $z$. The upper abscissa is scaled in units of the globally taken standard deviation of temperature, $T_{\mathrm{rms}}=\sqrt{\avg{T^2}_V}$.
  }
  \label{fig:pdf_T}
\end{figure}
\begin{figure}[!t]
  \centering
  {\footnotesize(a)}\hspace{0.5\textwidth}{\footnotesize(b)}\newline
  \includegraphics[width=0.53\textwidth]{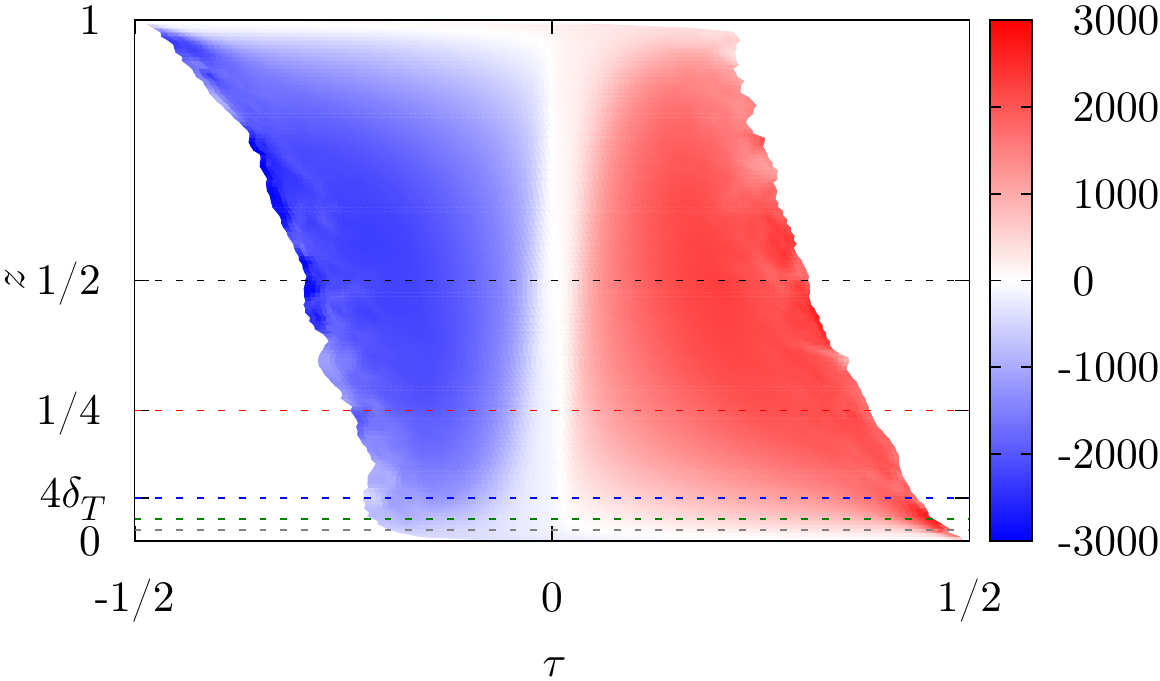}
  \hfill
  \includegraphics[width=0.45\textwidth]{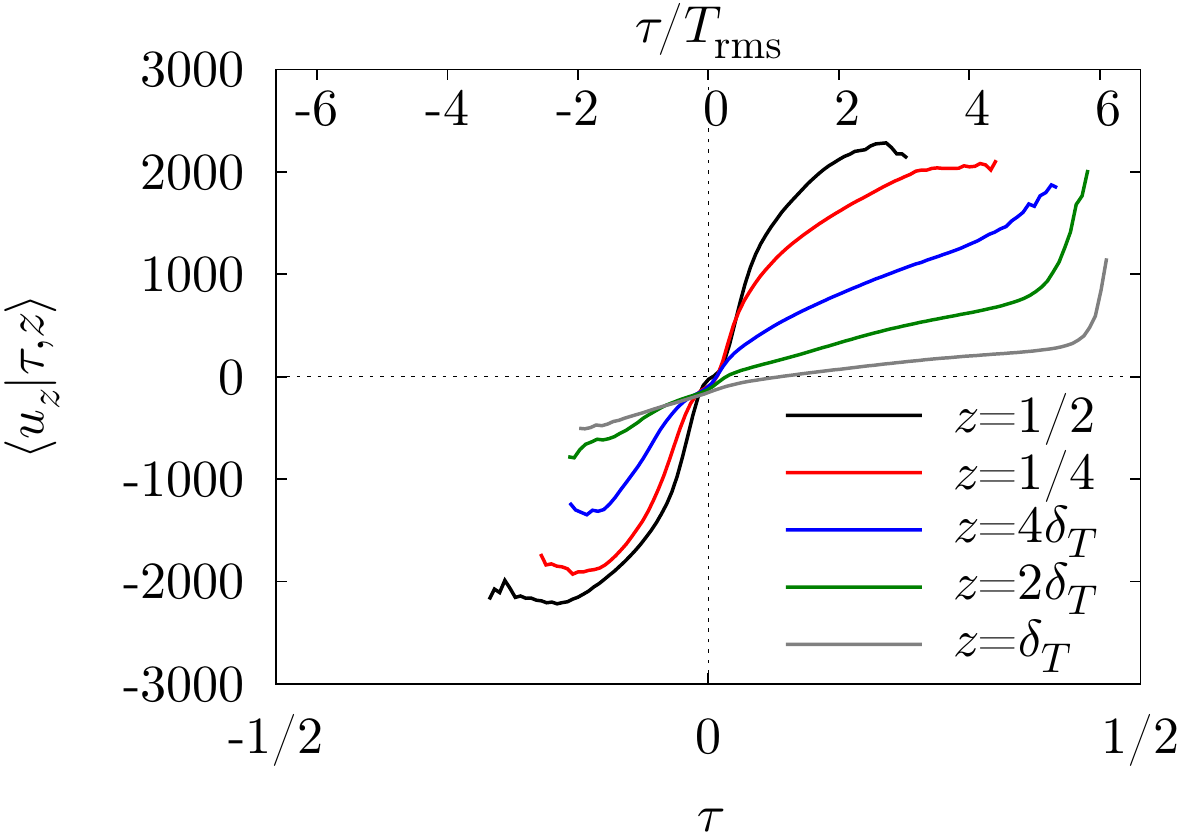}
  \caption{
    (a) Color plot of the conditional velocity field $\condavg{u_z}{\tau,z}$, with dashed lines as in \fref{fig:pdf_T}. Due to the non-dimensionalization, the velocity is given in units of the heat diffusion velocity, i.e. the velocity with which heat would be transported from plate to plate by pure heat diffusion.
    (b) The conditional velocity field $\condavg{u_z}{\tau,z}$ as a function of $\tau$ for various values of $z$.
  }
  \label{fig:condavg_uz_T}
\end{figure}

\Fref{fig:pdf_T}(a) shows the color-coded temperature PDF $h(\tau;z)$ in a logarithmic plot, where the black and gray lines indicate the mean temperature profile $\avg{T(z)}_A$ and the square root of the centralized second moment, $\Big\langle\big(T-\avg{T(z)}_A\big)^2\Big\rangle_A^{1/2}$, i.e.~standard deviation.
\Fref{fig:pdf_T}(b) shows slices in $\tau$-direction indicated by the dashed lines in \fref{fig:pdf_T}(a).
In the color plot, one clearly observes the sharp change of the temperature PDF from a $\delta$-function at the boundaries across the boundary layer to a shape exhibiting larger tails in the bulk.
In addition to these tails, another feature of the PDF is the hump close to the $\tau=0$--line.
This hump corresponds to the most probable value of the temperature.
One expects two different dynamical features to be responsible for this special shape; a tempting explanation would be to attribute the hump to the background temperature field of mean temperature, and account the wings for large $|\tau|$ for plumes that carry fluid that is much colder or hotter than the surrounding fluid.
An evidence for this is the shape of the PDF close to the bottom or top boundaries (but still outside the boundary layers):
At $z=4\delta_T$, though the most probable temperature value is moved slightly towards lower temperatures, the PDF exhibits a large tail at high temperatures.
The interpretation is that mostly cold fluid gathers in the lower regions of the bulk, being almost at rest (compare the region of $\condavg{u_z}{\tau,z=4\delta_T}$ in \fref{fig:condavg_uz_T}(b) corresponding to the hump), while very hot fluid is a more rare event, because hot fluid is convected away quickly due to plume dynamics.
The reason why the hot fluid takes greater temperature values than the cold fluid (in terms of absolute value) is that very cold fluid detaching from the top plate already heats up on its way down.

PDFs of the same shape for Rayleigh numbers of the same order are reported in \cite{emran08jfm}, where also the dependence on the vertical coordinate is taken into account.
The experimental data in \cite{castaing89jfm,ching93prl} shows a more pronounced exponential shape of the temperature PDF, which can be attributed to the difference in the Rayleigh numbers which are several orders of magnitude above ours; the numerical data in \cite{emran08jfm} suggests that the PDFs become more exponential with increasing Rayleigh number.
%Both these references and our findings report a strong dependence of the shape of the temperature PDF on the vertical position in the flow, which constrasts the theoretical derivation of an exponential PDF in all regions of the flow featured in \cite{yakhot89prl}.
% Both these references and our findings report a strong dependence of the shape of the temperature PDF on the vertical position in the flow, whereas a PDF that is exponential in all regions of the flow is derived in \cite{yakhot89prl}.
% This discrepancy arises due to the mean temperature profile being constant in the bulk of the flow (cf. \fref{fig:pdf_T}(a)), as pointed out in \cite{ching93prl}.

\begin{figure}[!t]
  \centering
  {\footnotesize(a)}\hspace{0.5\textwidth}{\footnotesize(b)}\newline
  \includegraphics[width=0.53\textwidth]{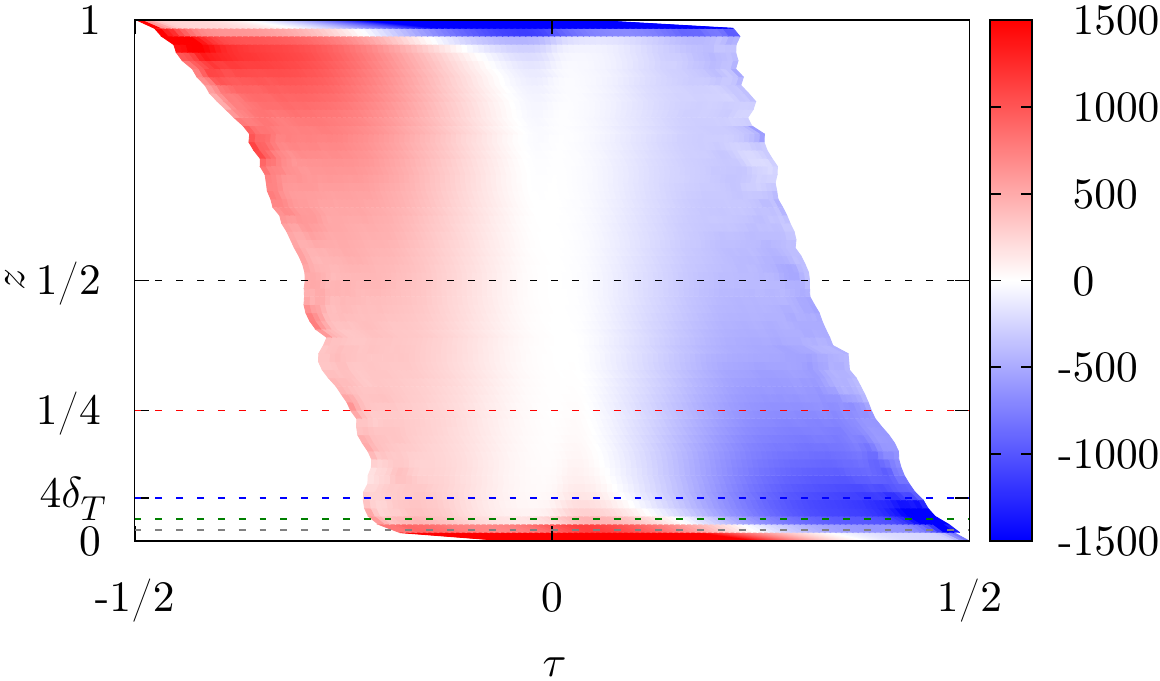}
  \hfill
  \includegraphics[width=0.45\textwidth]{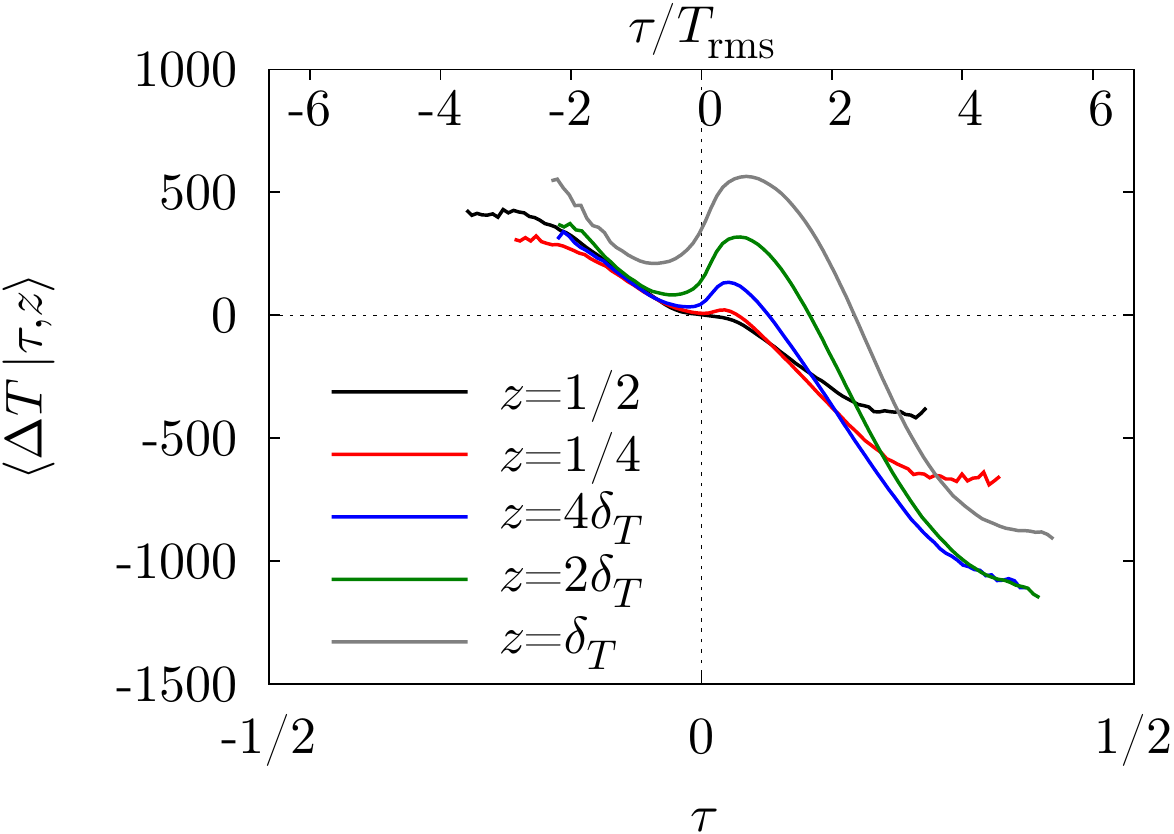}
  \caption{
    (a) Color plot of the conditional heat diffusion field $\condavg{\Delta T}{\tau,z}$, with dashed lines as in \fref{fig:pdf_T}.
    (b) The conditional heat diffusion field $\condavg{\Delta T}{\tau,z}$ as a function of $\tau$ for various values of $z$.
  }
  \label{fig:condavg_laplaceT_T}
\end{figure}
\Fref{fig:condavg_uz_T} and \ref{fig:condavg_laplaceT_T} exhibit the conditional averages introduced in \sref{sec:single_point_temperature_pdf}.
One can clearly observe the features that were suggested in the aforementioned section.
The conditional vertical velocity $\condavg{u_z}{\tau,z}$ is high (low) for hot (cold) fluid respectively, and the no-slip boundary conditions manifest in the fact that $\condavg{u_z}{\tau,z}$ is close to zero for $z\approx0$ and $z\approx1$.
Additionally, one observes a stripe close to the $\tau=0$--line of almost vanishing vertical velocity which coincides with the reddish core (the hump, i.e.~the most probable value) of the temperature PDF in \fref{fig:pdf_T}(a).
The interpretation is that fluid that is as hot as the mean temperature is neutrally buoyant and neither moves up nor down.
Another striking feature is the sudden increase of the vertical velocity for high $\tau$ near the boundary layer, i.e.~for $z=\delta_T$, which we attribute to rising plumes that detach from the hot bottom plate.
Again, it must be stressed that these interpretations hold in an averaged sense.

\Fref{fig:condavg_laplaceT_T} shows that the conditional heat diffusion term $\condavg{\Delta T}{\tau,z}$ is (in terms of absolute value) highest at the boundaries, with the term being positive (negative) at the hot bottom (cold top) plate.
On the contrary, in the bulk the absolute value is high (low) for very cold (hot) fluid, i.e.~in the wings of the temperature PDF.
Additionally, the $\tau$-slice near the boundary in \fref{fig:condavg_laplaceT_T}(b) shows an under- and overshoot.
The connection of these unique features to the RB dynamics has yet to be understood.

\begin{figure}[ht]
  \centering
  \includegraphics[width=0.85\textwidth]{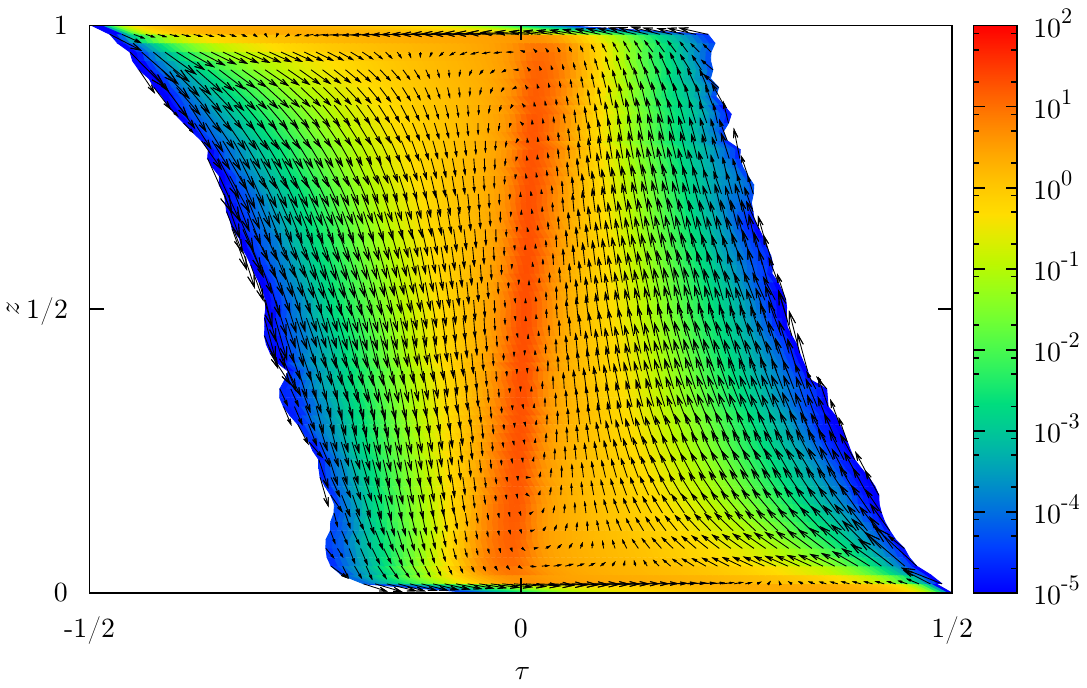}
  \caption{
    Color plot of the temperature PDF $h(\tau;z)$ together with the vector field identifying the characteristics.
  }
  \label{fig:condavg_laplaceT_uz_vectorfield}
\end{figure}
By combining the two aforementioned conditional averages to the vector field \eref{eq:characteristics} that defines the characteristics as suggested in \sref{sec:single_point_temperature_pdf}, one arrives at the vector field depicted in \fref{fig:condavg_laplaceT_uz_vectorfield} -- one of our central results.
It is easy to interpret this graph by tracing the vector field; one can qualitatively reconstruct the typical RB cycle of fluid heating up at the bottom, rising up while starting to cool down, cooling down drastically at the top plate, falling down towards the bottom plate while warming up a bit and heating up again at the bottom.
It is especially illustrative to see that the main contribution of cooling and heating (i.e.~biggest movement in $\tau$-direction of phase space) takes place near the boundaries, highlighting the importance of the boundary layers, while obviously the biggest movement in $z$-direction occurs in the bulk.

Yet one has to consider, for example, that although hot fluid rises up very quickly (referring to the vectors pointing upwards at the right side in \fref{fig:condavg_laplaceT_uz_vectorfield}), this does not contribute much to heat transport because these events occur rarely, as indicated by the temperature PDF shown along with the vector field governing the characteristics.

\begin{figure}[ht]
  \centering
  {\footnotesize(a)}\hspace{0.5\textwidth}{\footnotesize(b)}\newline
  \includegraphics[width=0.53\textwidth]{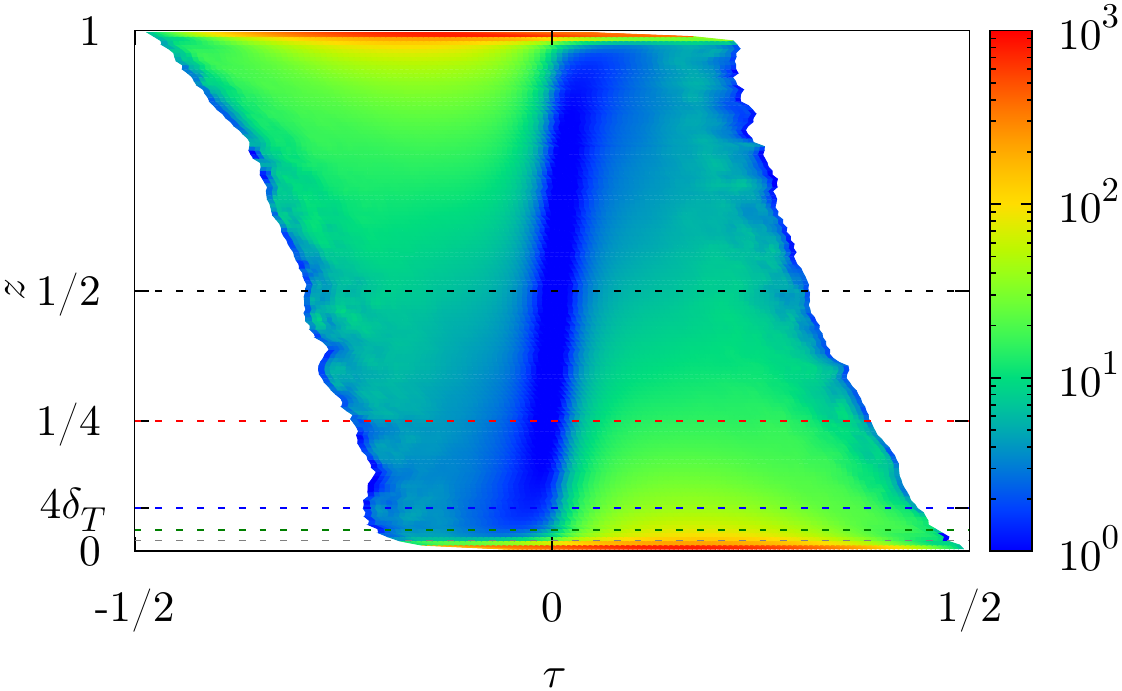}
  \hfill
  \includegraphics[width=0.45\textwidth]{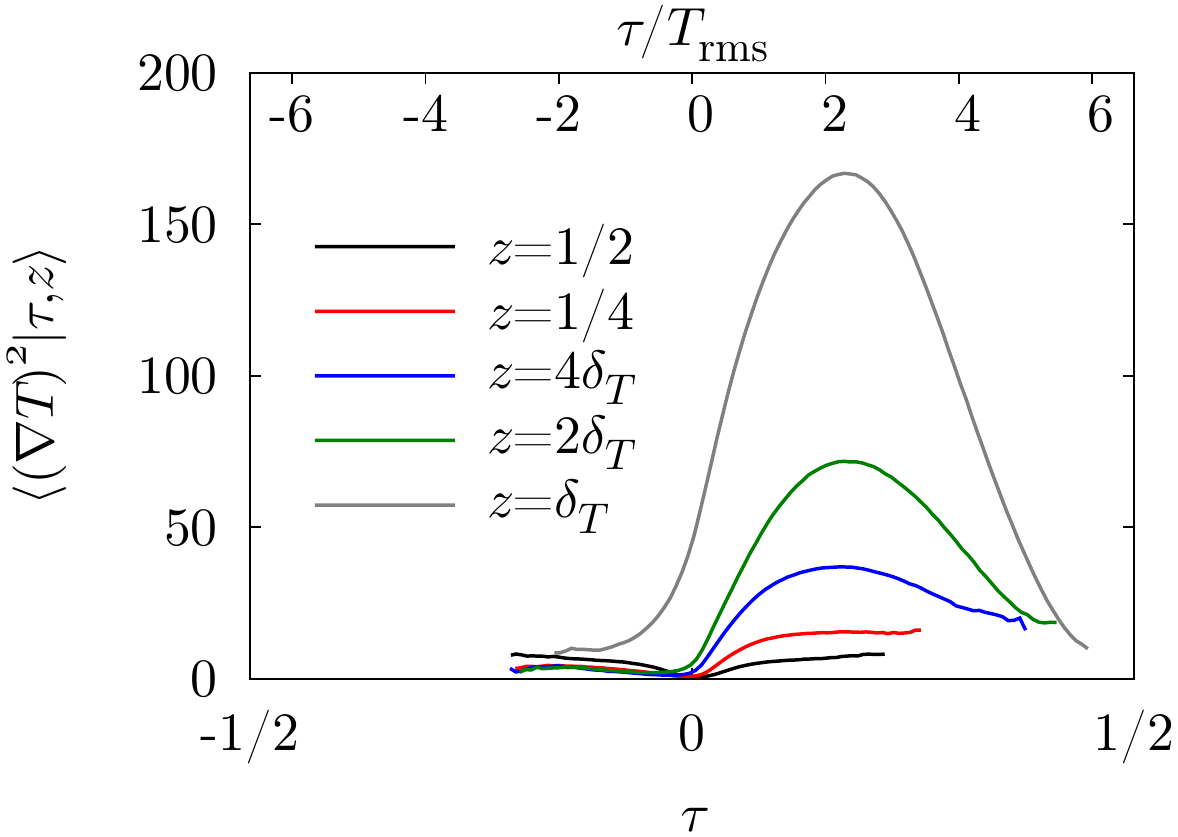}
  \caption{
    (a) Color plot of the conditional field $\condavg{(\nabla T)^2}{\tau,z}$, with dashed lines as in \fref{fig:pdf_T}. The color axis is scaled logarithmically.
    (b) The field $\condavg{(\nabla T)^2}{\tau,z}$ as a function of $\tau$ for various values of $z$.
  }
  \label{fig:condavg_nablaTsqr_T}
\end{figure}
In \fref{fig:condavg_nablaTsqr_T}, the conditional heat dissipation rate $\condavg{(\nabla T)^2}{\tau,z}$ is shown, which can be interpreted as a Nusselt number density according to \eref{eq:nusselt1}.
The noteable features are a pronounced minimum near the most probable temperature, again coinciding with the reddish core of the PDF.
Also, in the boundary layer this quantity attains huge values (note the logarithmic scaling in the color plot), which highlights the fact that the boundary layer contributes much to the heat transport.
A similar shape of a related quantity is reported in \cite{emran08jfm} (using the deviation from the mean temperature profile instead of the temperature itself), though there the conditional average is taken over the whole fluid volume and is hence lacking the $z$-dependency.

\section{Summary}\label{sec:summary}
In the present work, we have analyzed the single-point temperature PDF on the basis of the Lundgren-Monin-Novikov hierarchy by truncating the hierarchy on the first level via the introduction of conditional averages.

We have first derived the evolution equation of the full joint PDF of temperature and velocity.
Then we focused on the temperature PDF only, which is the central point of our paper, and obtained an evolution equation for it by reducing the joint PDF equation.
We assumed rather weak symmetry conditions of statistical stationarity in time and homogeneity in lateral spatial directions; these conditions should be fulfilled at reasonably high aspect ratios even for closed vessels, i.e.~are a good approximation of experimental setups in the bulk of the flow.
Under these symmetry considerations, the evolution equation of the temperature PDF becomes fairly simple.
The arising conditional averages of temperature diffusion $\condavg{\Delta T}{\tau,z}$ and vertical velocity $\condavg{u_z}{\tau,z}$ are estimated by direct numerical simulations using a suitably designed penalization approach, and features of them are discussed.
It shows that expected features such as properties of the temperature and velocity boundary layers, correlation of temperature and velocity and so on are related to the form of these conditional averages that naturally come up in our derivations.

The evolution equation of the temperature PDF is readily treated by the method of characteristics.
Due to the applied symmetry conditions, the phase space which describes our system becomes two-dimensional, spanned by temperature $\tau$ and vertical coordinate $z$.
Because of this reduced dimensionality of the system, the method of characteristics yields a descriptive view of the RB dynamics, resulting in the vector field describing the evolution in $\tau$-$z$--phase space.
The characteristics, i.e.~trajectories in $\tau$-$z$--phase space, are found to reproduce the typical cycle of a fluid parcel.
The regions of the main transport in $\tau$-direction have been identified as the boundary layers, while the major movement in $z$-direction takes place in the bulk.
This highlights the importance of the boundary layers to the heat transport.
The relation of heat transport in terms of the Nusselt number to the conditional averages introduced in our derivation is briefly discussed, leading us to the definition of a Nusselt number density.
It would be very interesting to obtain the statistical quantities describing the evolution of the PDF directly from experiments, e.g.~from measurements of instrumented particles as described in \cite{gasteuil07prl}.

Future efforts will be to not only use the characteristics as an illustrative way to describe the mean movement in phase space, but actually calculate the PDF of temperature from the integral representation \eref{eq:temperature_pdf_integral}.
Also, modeling of the conditional averages, which are up to now estimated from direct numerical simulations, might be feasible; an intermediate step would be to discuss the quantities not in the turbulent case, but close above the bifurcation from heat conduction to convection, where analytical solutions of temperature and velocity fields are available.
Though an easy illustration in the form of trajectories in two-dimensional space will not be achievable in the case of the joint PDF, this approach is nevertheless promising and planned for the future, because we hope that already the form of the conditional averages will give insight into the connection of the statistics to the RB dynamics.
An intermediate step would be to concentrate on the joint PDF of temperature and vertical velocity, which should among others relate to the dynamics of plumes.

\section*{References}
\bibliographystyle{unsrt}
\bibliography{references}

\end{document}